\documentclass[amsmath,amssymb,twocolumn,pre,floatfix]{revtex4}
\usepackage{graphicx}
\usepackage{epstopdf}
\def\D{{\rm d}}
\def\E{{\rm e}}
\def\I{{\rm i}}
\newcommand{\tf}[1]{\textstyle\frac #1}
\begin{document}
\title{Vector Formalism for Active Nematics in Two Dimensions }
\author{L.~M. Pismen}
\affiliation{Department of Chemical Engineering, Technion -- Israel Institute of Technology, Haifa 32000, Israel}
\begin{abstract}
Specific features of two-dimensional nematodynamics give rise to shortfalls of the tensor representation of the nematic order parameter commonly used in computations, especially in theory of active matter. The alternative  representation in terms of the vector order parameter follows with small adjustments the classical director-based theory, but is applicable to 2D problems where both nematic alignment and deviation from the isotropic state are variable. Stability analysis of nematic alignment and flow is used as a testing ground. A director-based analysis demonstrates a shortfall of the standard theory, which does not ensure relaxation to equilibrium in a passive system. It also demonstrates the inadequacy of the director-based description, which misses a stabilizing effect of perturbations of the modulus ensuring stability of a passive system on scales far exceeding the healing length.
\end{abstract}
 \maketitle

\section{Introduction} 

Studies of active nematics commonly deal with 2D patterns in thin layers with tangential alignment on containing walls \cite{yom}, as well as in thin elastic sheets \cite{andr} and cellular layers \cite{cell1,cell2}. Tensor representation is commonly viewed as necessary when not only the orientation but the modulus of the nematic director is variable. The tensor description has been pioneered by Doi \cite{doi} and extended to include spatial inhomogeneities \cite{eb89} and interactions between the nematic alignment and flow \cite{eb94,qs98,sl05}. The tensor order parameter $ Q_{ij}=\varrho (n_in_j-\delta_{ij}/d)$ contains the modulus $\varrho$ and is made traceless by adding the last term, where $d$ is the number of dimensions. The immediate advantage of the tensor order parameter over the director \textbf{n} is, besides the presence of the modulus, its invariance to rotation by $\pi$ stemming from its dependence on twice the inclination angle. 

Even though computational models are commonly restricted to 2D,  they are often based on the Beris--Edwards 3D model \cite{eb94} adjusted to 2D just by setting $d=2$ in the above definition. In this way, one does not take advantage of substantial simplifications brought about by a lower dimensionality. Moreover, as argued below, a simpler approach based on a vector order parameter not only allows for full description of the nematic state in 2D, but has a preference over the tensor-based theory by avoiding the loss of distinction between splay and bend elasticity. 

\section{Specifics of two dimensions} 

\subsection{A missing elastic constant} 

The tensor-based Landau--de Gennes Lagrangian (LdGL) in 2D is
\begin{equation} 
 \mathcal{L} = -\alpha_0[ Q_{ij}Q_{ij}+\tf12(Q_{ij}Q_{ij})^2 ]+ \mathcal{L}_\mathrm{d},
  \label{elastK3} \end{equation}
where the coefficient at the fourth-order term is chosen to set the value of the modulus $\varrho$ to unity in a perfectly aligned nematic; summation over repeated indices is implied throughout. The two constituents of the distortion Lagrangian  $\mathcal{L}_\mathrm{d}$ are
\begin{align}
\mathcal{L}_1&=\tf12K_1|\partial_kQ_{ij}|^2 = \tf14K_1 (|\nabla q_1|^2 +|\nabla q_2|^2), \cr
 \mathcal{L}_2 &=\tf12K_2|\partial_jQ_{ij}|^2    \label{2QL}\\ &=
 \tf14K_2 [ (\partial_1 q_1+\partial_2 q_2)^2+ (\partial_2 q_1-\partial_1 q_2)^2].
\notag  \end{align}
Another possible construction, $|\varepsilon_{jk}\partial_kQ_{ij}|^2$, where $\boldsymbol{\varepsilon}$ is the antisymmetric tensor, leads to the expression identical to $\mathcal{L}_1$. Varying the distortion energy  $\mathcal{F}_\mathrm{d}=\int  \mathcal{L}_\mathrm{d} \D \mathbf{x}$ defines the molecular field tensor
\begin{equation} 
h_{ij}=-\partial\mathcal{L}/ \partial Q_{ij} +  \partial_k \pi_{ijk} , \quad
  \pi_{ijk}=\partial\mathcal{L}/ \partial(\partial_kQ_{ij}).  
 \label{elasth3} \end{equation}
The tensor \textbf{Q} is conveniently parametrized as 
\begin{equation}
\mathbf{Q}=\tf12\begin{pmatrix} q_1 & q_2 \\  q_2 & -q_1\end{pmatrix}, \quad
\begin{array}{r}  q_1=\varrho \cos 2\theta,  \\ \quad q_2=\varrho \sin 2\theta, \end{array}
 \label{elastQ} \end{equation}
where $\theta$ is the inclination angle and $\varrho=\sqrt{ q_1^2+ q_2^2}$. The molecular field is a vector when expressed through the vector \textbf{q} but still based on the tensor LdGL
\begin{equation} 
h_k =- \partial \mathcal{L}/\partial q_k + \partial_i \pi_{ik},\quad \pi_{ik} =\partial \mathcal{L}_\mathrm{d}/\partial(\partial_iq_k).
 \label{elasth2} \end{equation}

 Two independent constants are sufficient in 2D, but extra terms in $\mathcal{L}_2$, $\partial_1 q_1\partial_2 q_2-\partial_2 q_1\partial_1 q_2$ vanish upon variation, so that they do not contribute to the molecular field, and its distortion part is expressed as $h_i^\mathrm{d}=\frac12 K\nabla^2q_i$ with $K=\frac12(K_1+K_2)$ or, equivalently,  $h_{ij}^\mathrm{d}=K\nabla^2Q_{ij}$. Similarly, in 3D there are only two independent elastic constants in the leading order. A third constant could be only obtained by adding a higher-order term $Q_{ij}\partial_iQ_{kl}\partial_jQ_{kl}$. It does not look contentious that different terms in the classical director-base expression of the distortion energy, $( \mathrm{div}\, \mathbf{n})^2, \ ( \mathbf{n} \cdot \mathrm{curl}\, \mathbf{n})^2, \, ( \mathbf{n} \cdot \mathrm{curl}\, \mathbf{n})^2$ contain different powers of the \textbf{n}. But the tensor order parameter, unlike \textbf{n}, has a physical reality, and it is an apparent drawback that elastic anisotropies cannot be fully captured in the framework of a standard Landau model restricted to the fourth combined order in \textbf{Q} and spatial derivatives. 

Most commonly, the ``one-constant approximation" is used in simulations, assuming all elastic coefficients to be equal. Since LdGL is, in any case, a crude approximation allowing for qualitative insights, equality of bend and splay elasticities may appear to be not quite disturbing: the distinction would be certainly restored in higher orders. However, this turns out to be an artifact of the tensor formalism in 2D. 

\subsection{Vector Nematodynamics}

The tensor \textbf{Q} can be constructed by merging two vectors, $\mathbf{q}/\sqrt{2}$ and $\mathbf{q}'/\sqrt{2}$, where $q'_i=\varepsilon_{ij}q_j$. Hence, the vector $\mathbf{q}$, which is  invariant under rotation by $\pi$, alone carries all necessary information, but the Q-tensor formed by this merger can be brought back whenever a linear tensor expression is cslled for. It might be disturbing that $\mathbf{q}$, as a vector, is odd under parity inversion, but we shall see that it does not affect the relations derived below. The important advantage of the vector-based description is restoring the distinction between bend and splay elasticities. In 2D, $\mathrm{curl}\, \mathbf{q}$ is a pseudoscalar, and LdGL is expressed, without involving higher orders of $\mathbf{q}$, as 
\begin{equation} 
 \mathcal{L} =-\tf12\alpha_0|\mathbf{q}|^2 \left(1-\tf12|\mathbf{q}|^2\right) + \tf12 K_1( \mathrm{div}\, \mathbf{q})^2 
 + \tf12 K_2 ( \mathrm{curl}\, \mathbf{q})^2,  
 \label{elast2q} \end{equation}
where, for the sake of similarity to the director-based description, the elasticities $K_i$ are multiplied by two. This leads to the molecular field
\begin{align} 
\mathbf{h}&=\alpha_0\mathbf{q}(1-\varrho^2) +\mathbf{h}^\mathrm{d}, \quad  h_k^\mathrm{d}=\partial_i\pi_{ik}, \cr
 h_1^\mathrm{d} &=  K_1 (\partial^2_1  q_1 +\partial_1\partial_2  q_2) 
 +K_2(\partial^2_2  q_1 -\partial_1\partial_2  q_2), \cr 
  h_2^\mathrm{d} &=  K_1 (\partial^2_2  q_2 +\partial_1\partial_2  q_1) 
  + K_2(\partial^2_1  q_2 -\partial_1\partial_2  q_1).
 \label{2pih} \end{align}
or, setting $K=\tf12(K_1+K_2), \, K'=\tf12(K_1-K_2)$,
\begin{align} 
 h_1^\mathrm{d} &= K \nabla^2q_1 +K'[(\partial^2_1-\partial^2_2  )q_1 +2\partial_1\partial_2  q_2], \cr 
  h_2^\mathrm{d} &=  K \nabla^2 q_2 +K'[(\partial^2_2- \partial^2_1 )q_2 +2\partial_1\partial_2  q_1].
 \label{2pih2} \end{align}
The simple expression $\mathbf{h}^\mathrm{d}= K\nabla^2\mathbf{q}$ is obtained at $K'=0$, but generally this constant does not vanish. This kind of rank reduction is specific to 2D, since the bend and twist terms in 3D would require a higher power of \textbf{q}, but it testifies of an advantage of the vector, compared to tensor, formalism in 2D. 

Constructing the Lagrangian based on the vector \textbf{q} in leu of the Q-tensor enables extending  to 2D systems with a variable modulus the classical Ericksen--Leslie (EL) formalism \cite{dgp} with minimal adjustments. As in the standard director-based theory, the acceleration of the fluid is determined by the generalized Navier--Stokes equation
\begin{equation}
 \rho D_t v_i = \partial_j ( \sigma^\mathrm{s}_{ij} +\sigma_{ij}^\mathrm{d}- p\delta_{ij})
\label{nsg}\end{equation}
where $D_t=\partial_t + \mathbf{v}\cdot \nabla $ denotes the substantial derivative accounting for advection with flow velocity $\mathbf{v}$, $\boldsymbol{\sigma}^\mathrm{s}$ is the viscous stress tensor, $\sigma_{ij}^\mathrm{d}= -\pi_{ik}\partial_jq_k$ is the distortion stress, $\rho$ is density, and $p$ is pressure. The dynamics of the order parameter \textbf{q} stems from the change of entropy $S$ at a constant temperature $T$, which is is expressed in an incompressible fluid as
\begin{equation}
TD_t S= \int \left(\sigma_{ij}^\mathrm{s} s_{ij}+ \mathbf{h\cdot N}-p \nabla \cdot \mathbf{v}\right) \D\mathbf{x},
\label{entr1}\end{equation}
where pressure serves as a Lagrange multiplier enforcing the continuity equation $\nabla \cdot \mathbf{v}=0$ and $s_{ij}$ is the shear tensor, further separated into the symmetric and antisymmetric parts:
\begin{equation}
s_{ij}=\tf12 (\partial_iv_j +\partial_jv_i), \quad  
\Omega_{ij}=\frac12 (\partial_iv_j -\partial_jv_i) =\tf 12\varepsilon_{ij} \omega,
\label{symasym}\end{equation}
where $\omega=\nabla\times \mathbf{v}$ is the vorticity pseudoscalar. The vector $\mathbf{N}$ represents the rate of the advection and rotation of the order parameter by fluid flow. At this point, a slight modification is due. Rotation does not affect the modulus, and the distortion energy does not change when both the nematic alignment and the fluid as a whole are rotated with the same angular velocity. Since \textbf{q} rotates with twice director's speed, the director-based energy-conserving counterpart to a rotation $\partial_t x_i=\omega\varepsilon_{ij}x_j$ is, rather than $\partial_t n_i= \omega\varepsilon_{ij}q_j$ in the EL formalism, is
\begin{equation}
\partial_t q_i= \tf12\omega\varepsilon_{ij}q_j,
 \quad \partial_t x_i=\omega\varepsilon_{ij}x_j.  
\label{rotom}\end{equation}
Accordingly, the antisymmetric part of $\boldsymbol{\sigma}^\mathrm{s}$ is halved to $\tf14 \mathbf{q\times h}$. We also take note that, since the 2D vorticity is a pseudoscalar, the expression $\boldsymbol{\omega} \times \mathbf{q}$ makes no sense in 2D, and the change of \textbf{q} due to advection and rotation is presented by vector $\mathbf{N}=D_t \mathbf{q} -\tf12 \boldsymbol{\Omega}\cdot\mathbf{q}$. 

The contributions to the entropy source are presented as products of fluxes and conjugate forces with coefficients having the dimension of viscosity. Based on \eqref{entr1}, $\sigma_{ij}^\mathrm{s}$ is identified as the force conjugate to $s_{ij}$ and $\mathbf{h}$ as  the force conjugate to $\mathbf{N}$. The forces are tied to the fluxes by linear relations with coefficients having the dimension of viscosity arranged into tensors of an appropriate rank. In the EL theory, higher-rank tensors accounting for viscous anisotropies are composed by adding higher powers of \textbf{n} with respective orientational viscosities. A similar construction with higher powers of \textbf{q} adds higher powers of the modulus. The lowest-order expression is
\begin{align}
 \sigma_{ij}^\mathrm{s}&=\mu s_{ij}+ \tf12 \gamma'_2 (q_iN_j+q_jN_i), \label{sdyn} \\ 
 h_i&=\gamma_2 q_j s_{ij} + \gamma_1N_i,
   \label{hdyn}\end{align}
where $\mu$ is the common dynamic viscosity. The Onsager reciprocity relations determine the relation  between the parameters $\gamma'_2=\gamma_2$. The dynamic equations of the vector \textbf{q}, which is the counterpart of the  director-based equation
\begin{equation}
D_t n_i =\Omega_{ij}n_j -\chi s_{ij}q_j+\Gamma h_i,
\label{ndyn}\end{equation}
follows from \eqref{hdyn}:
\begin{equation}
D_t q_i =\tf 12\Omega_{ij}q_j -\chi s_{ij}q_j+\Gamma h_i.
\label{qdyn}\end{equation}
Here $\Gamma=1/\gamma_1$ is the mobility parameter and $\chi= \gamma_2/ \gamma_1$ is the dimensionless alignment parameter, which determines the response of the director to local shear. The system \eqref{nsg}, \eqref{hdyn}, \eqref{qdyn} describes the combined evolution of the order parameter and flow. In the theory of active dynamics, the sum $\boldsymbol{\sigma}^\mathrm{s}+\boldsymbol{\sigma}^\mathrm{d}$ is called passive stress, expressed when based on $\mathbf{n}$ or $\mathbf{q}$ as
\begin{align}
  \sigma^\mathrm{p}_{ij}&=\tf 12 \chi (n_ih_j+h_in_j)-\tf 12 (n_ih_j-h_in_j) +\pi_{ik}\partial_jn_k , \label{sign}\\
  \sigma^\mathrm{p}_{ij}&=\tf 12 \chi (q_ih_j+h_iq_j)-\tf 14 (q_ih_j-h_iq_j) +\pi_{ik}\partial_jq_k . 
\label{sigp}\end{align}

\section{Stability Analysis}

\subsection{Instabilities in the director-based theory\label{s3a}}

We will further apply the vector-based formalism to the analysis of instabilities caused by the feedback loop between advection-induced distortions of a nematic alignment field and flow it generates \cite{simram}. In order to emphasize the role of perturbation of the modulus, we start with the analysis in the framework of the EL theory, elaborating upon the numerical study in a confined geometry \cite{edyom}.

Onsager reciprocity relations do not apply to active media, so that, strictly speaking, the entire nematodynamic theory based on these relations should be revised. Short of going this far, theorists commonly add the active stress $\sigma^\mathrm{a}$, which is expressed through the director as $\sigma^\mathrm{a}_{ij}=-\zeta n_in_j$ with the parameter $\zeta$ positive for extensile and negative for contractile activity. This expression, taken literally, is problematic, since entropy-changing contributions to the stress tensor have to be odd under time reversal \cite{bp14}. The contradiction can be resolved by letting $\zeta=\D_tc$, where $c$ is the concentration of an active chemical, such as ATP in intracellular processes. Then the term becomes odd  under time reversal, as required, and can shift entropy either way. If there is no feedback of alignment on the active chemical and its consumption rate is constant, chemistry may remain behind the scenes.  

We first test a passive system, where a perfectly ordered state is expected to be stable. However, even in this case, there are two questions to decide: whether perturbations of a perfectly aligned state may cause a deviation of the modulus from unity and whether arbitrary values of the alignment parameter are admissible. We consider perturbations of an ordered state with $|\mathbf{n}| =1$ in an unbounded domain. In view of the circular symmetry with respect to the unperturbed alignment, it is sufficient to view the problem in 2D: this is the case when free-flow description is applicable. A constant background velocity is irrelevant due to Galilean invariance, and can be eliminated by transforming to a comoving frame. The unperturbed alignment direction is irrelevant in an unbounded domain, and can be chosen along the $x_1$ coordinate, so that $n_1^0=1, \,n_2^0=0$. 

In the leading order, only a small perturbation $\epsilon n_2$ does not perturb the modulus. We expand it in the Fourier series $n_2 = \epsilon \widehat{n}_2 \E^{\I\mathbf{ k\cdot x}}$ where $\mathbf{k}=k\{\cos \varphi, \sin \varphi\}$ is the wave vector. The molecular field, defined by the distortion part of \eqref{2pih} with $\mathbf{q}$ replaced by $\mathbf{n}$, is transformed as 
\begin{align}
\widehat{h}_1&=-\tf12 k^2 \widehat{n}_2(K_1-K_2)\sin 2\varphi, \cr 
\widehat{h}_2&=- k^2 \widehat{n}_2 (K_1\sin^2\varphi+K_2\cos^2\varphi).
 \label{passf} \end{align}
The distortion stress vanishes upon linearization, and the passive stress \eqref{sign} is linearized and transformed as 
\begin{align}
\widehat{\sigma}^\mathrm{p}_{11}&=\tf12 \chi \widehat{h}_1, \quad,
\widehat{\sigma}^\mathrm{p}_{12}=\tf12 ( \chi +1)\widehat{h}_2, \cr
\widehat{\sigma}^\mathrm{p}_{21}&=\tf12 ( \chi-1)\widehat{h}_2, \quad \widehat{\sigma}^\mathrm{p}_{22}=0.
\label{passf}\end{align}

It is advantageous to express the perturbation velocity $\widetilde{\mathbf{v}}$ through the stream function, $\widetilde{v}_i =\varepsilon_{ij}\partial_j\Psi$, expand the latter as $\Psi=\epsilon\widehat{\Psi}\E^{\I\mathbf{ k\cdot x}}$, and take the curl of \eqref{nsg}. The contribution of passive stress is expressed then as $\varepsilon_{ik}\partial_k\partial_j\widehat{\sigma}^\mathrm{p}_{ij}$, which leads to the Fourier-transformed hydrodynamic equation
\begin{align}
\left(\rho D_t  +\mu  k^2\right) \widehat{\Psi}& =- \tf12 k^2 \widehat{n}_2 [ (\chi+1)K_1\sin^2\varphi \cr 
 &+ (\chi-1)K_2\cos^2\varphi].
\label{passpsi}\end{align}
The transformed linearized equation \eqref{ndyn} also defines dynamics of $\widehat{n}_1$: 
\begin{align}
D_t \widehat{n}_1 &= \Gamma\widehat{h}_1 +\tf 12\chi \widehat{\Psi}\sin 2\varphi, \cr
D_t \widehat{n}_2 &= \Gamma \widehat{h}_2  -\tf 12\widehat{\Psi}(1+\cos 2\varphi).
\label{ndynt}\end{align}
Any non-zero $n_1$ causes deviation of the modulus from unity. However, the equations of the other two variables do not depend on $\widehat{n}_1$. Therefore the director-based analysis is still justified as long as the alignment is stable, but an instability excites $n_1$ and, hence, changes the modulus as well. 

Stability is determined by the eigenvalues $\lambda$ of the matrix \textbf{J} defining the dynamics of $\widehat{\Psi},\,\widehat{n}_2$. Its elements are
\begin{align}
J_{\Psi\Psi}&=- \mu , \quad J_{n\Psi}= -\tf 12 (1+\chi \cos 2\varphi), \cr 
J_{\Psi n}&=- \tf12 [ (\chi+1)K_1\sin^2\varphi + (\chi-1)K_2\cos^2\varphi] , \cr
J_{nn}&=- \Gamma(K_1\sin^2\varphi+K_2\cos^2\varphi)] .
\label{Ldyn} \end{align}
Common powers of $k$ are cancelled here in each row, so that if instability arises, it happens simultaneously on all wavelengths. The trace of  \textbf{J} is always negative, so that an oscillatory instability is excluded, and the determinant is
\begin{align}
& K_1\sin^2\varphi[\mu \Gamma-\tf12(\chi+1) (1+\chi \cos 2\varphi) ]\cr
+&K_2\cos^2\varphi[ \mu \Gamma-\tf12 (\chi-1) (1+\chi \cos 2\varphi) ].
\label{jdet}  \end{align}
It turns out that stability is not warranted in the entire parametric domain. This can be seen most clearly at $K_1=K_2=K$ when the determinant simplifies to 
\begin{equation}
 K[ \mu \Gamma + \tf 14\chi ^2(1+\cos 4\varphi) -\tf 12].
    \label{jdet0}\end{equation}
The part of the determinant excluding $\mu \Gamma$ is negative in the flow-tumbling regime at $|\chi|<1$ when the director continuously rotates under shear, but also at $|\chi|>1$, when the director tends to align at a certain angle to the flow direction, negative values persist at $|\chi|<2$ at $\varphi$ close to $\pm \pi/2$. A perfectly aligned stationary state is always stable only at $\Gamma\mu>2$. This should be impossible in a fully dissipative system, but is not ruled out since substituting the dynamic equations of the director and stream function into the equation of entropy change \eqref{entr1} mixes variables odd and even with respect to time reversal \cite{bp}. This computation confirms that the standard theory \cite{dgp} does not ensure relaxation to equilibrium. It also demonstrates inadequacy of the director-based description: we shall see below that perturbations of the modulus serve as an efficient stabilizing factor. 

Active stress adds to the hydrodynamic equation the term $\tf12k^2\zeta \widehat{n}_2\cos 2\varphi$ containing a lower power of $k$, so that it dominates the determinant at long scales, and an arbitrarily small activity causes instability at some perturbation angle $\varphi$. The leading term of the determinant flips its sign when $\cos 2\varphi$ equals zero or $-\chi^{-1}$, so that both extensile and contractile activity causes instability at some perturbation angle. This is corrected in a realistic case of a confined geometry. 
 
\begin{figure}[t]\centering
\includegraphics[width=0.4\textwidth]{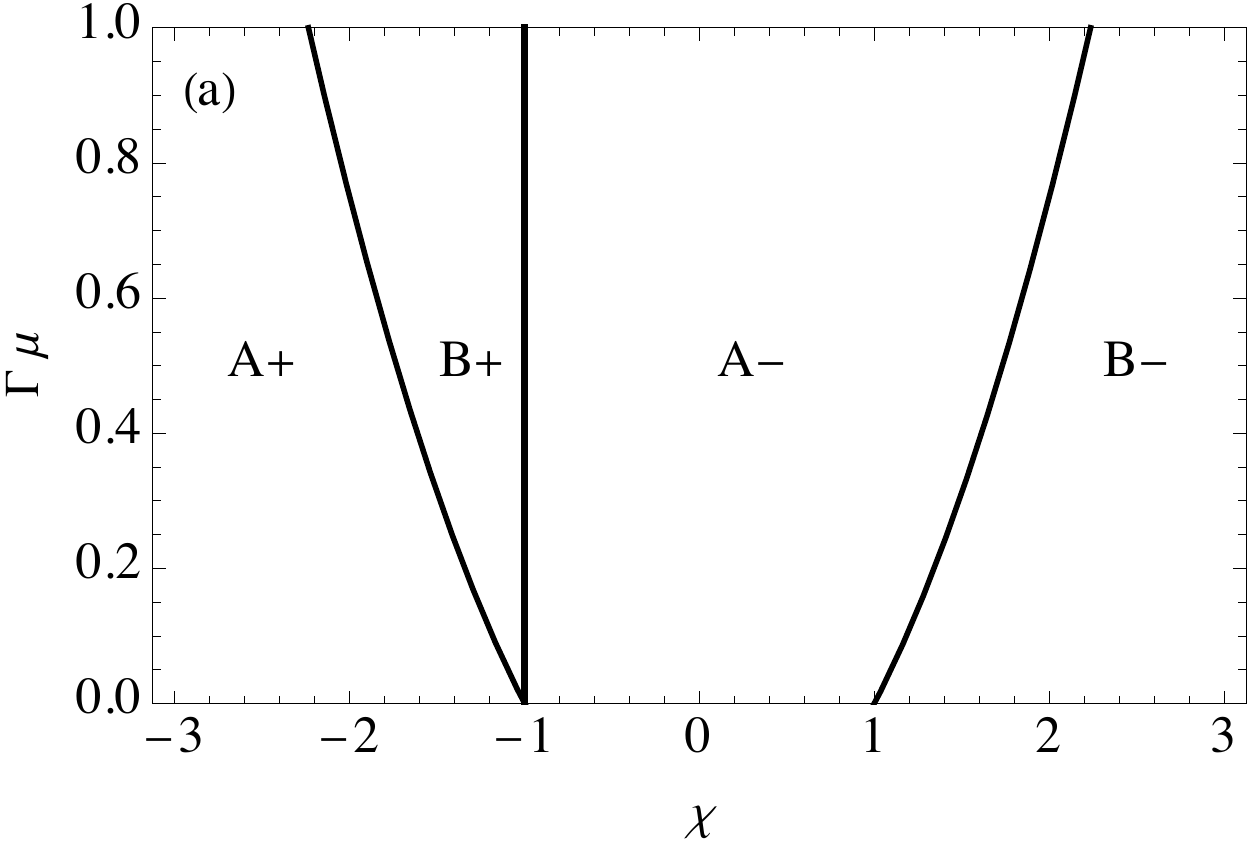} \\ \includegraphics[width=0.4\textwidth]{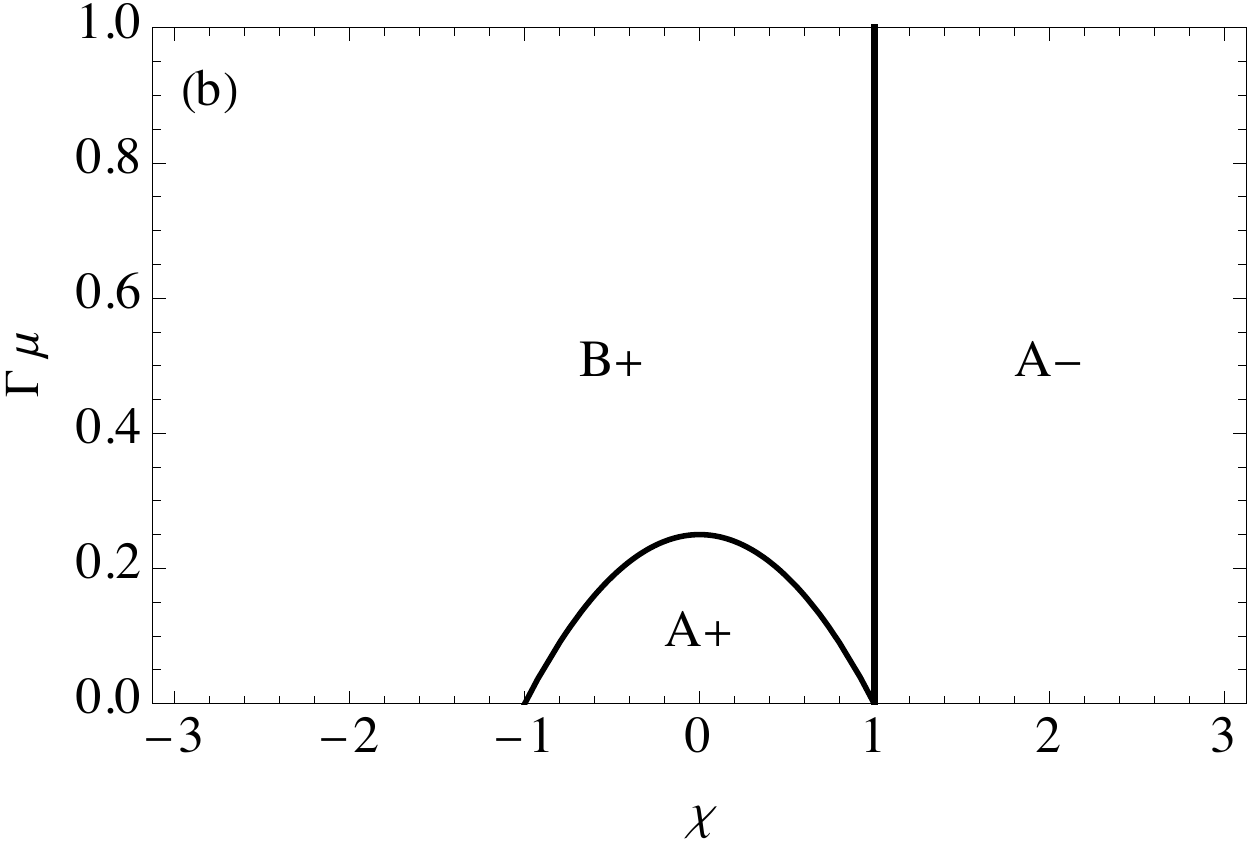}
\caption{Instability regions on a cylinder wit the axis parallel (a) and perpendicular (b) to the unperturbed orientation. See the text for explanations.}
\label{fns}\end{figure}

A geometry allowing for perturbations at any angle is the surface of a cylinder with the radius $l$. Then the largest available wavenumber is $k=l^{-1}$ at $\varphi$ equal to the angle between the axis of the cylinder and the unperturbed orientation of the director, and the instability threshold is    
\begin{eqnarray}
 \zeta &=&\frac{l^{-2}}{ \cos 2\varphi}\bigg[ 
 \frac{4\mu\Gamma(K_1\sin^2\varphi+K_2\cos^2\varphi)}{1+\chi\cos 2\varphi}\cr 
&+& K_1(\chi+1)\sin^2\varphi+K_2(\chi-1)\cos^2\varphi\bigg].
\label{sp1}\end{eqnarray}
If the cylinder is wrapped along the $x_2$ or $x_1$ coordinate, the largest wavenumber is at $\varphi=0,\, \pi$ or $\varphi=\pm\pi/2$, respectively. In the first case, the instability threshold $\zeta^*$ solves 
\begin{equation}
l^2 \zeta^* (\chi+1) +K_2(4\mu\Gamma+1-\chi^2)  =0.
\label{sp0}\end{equation}
The threshold diverges at $\chi=-1$ and vanishes at $\mu\Gamma=\tf14 (\chi^2-1)$. At $\chi>-1$, contractile activity is destabilizing; it should exceed a negative value of $\zeta^*$ by its absolute value when $\mu\Gamma>\tf14 (\chi^2-1)$ (region $A-$ in  Fig.~\ref{fns}a), but at lower values of  $\mu\Gamma$ (region $B-$), an arbitrary small contractile activity is destabilizing. This is mirrored at $\chi<-1$ with extensile activity destabilizing when it is sufficiently strong  at 
$\mu\Gamma<\tf14 (\chi^2-1)$ (region $A+)$ and at an arbitrary strength otherwise (region $B+$). In the second case,  
$\zeta^*$ solves
\begin{equation}
 l^2 \zeta^* (\chi-1) = - K_1(4\mu\Gamma-1+\chi^2).
\label{sppi}\end{equation}
The regions $A\pm$ and $B+$ are situated as in Fig.~\ref{fns}b, while the region $B-$ is missing.

More constraints appear in a domain bounded by solid walls with no-slip boundary conditions, but the situation when instability may be caused by an arbitrary small activity is nonphysical. It reflects the appearance of instabilities in the absence of activity and reveals once more the inadequacy of the director-based description.

\subsection{Instabilities in the vector-based theory\label{s3b}}

Adjusting the above analysis to the case os a variable modulus, we start again with testing a passive system and consider perturbations of the ordered state $q_1^0=1, \,q_2^0=0$ in an unbounded domain (which may be, as before, a 2D slice of  3D space). The description allowing for a variable modulus has an intrinsic scale of energy density $\alpha_0$ and the length scale $\xi= (K_1/\alpha_0)^{1/2}$, which is a microscopic healing length. The natural time scale is $(\alpha_0\Gamma)^{-1}$. The  dimensionless parameters are $\widehat{\mu}=\mu\Gamma\xi, \, \widehat{\rho}=\rho \alpha_0\Gamma^2\xi^3$. We will further use them with hats omitted, and retain the same notation for the variables in dimensionless equations. 

Since the circular symmetry allows for a 2D description even in the 3D world, it is possible to neglect friction for the time being in order to clearly see the effect of a variable modulus. As before, deviations from the base state $\epsilon \widetilde{q}_i$ are expanded in the Fourier series $\widetilde{\mathbf{q}} = \epsilon \widehat{\mathbf{q}} \E^{\I\mathbf{ k\cdot x}}$. Unlike \eqref{passf}, the molecular field \eqref{2pih} has an algebraic part in the dimensionless form $q_i(1-\varrho^2)$, reduced now to a single component $\widehat{\mathbf{h}}^0_1=-2\epsilon\widehat{q}_1$. The distortion part of \eqref{2pih} is transformed as $\widehat{\mathbf{h}}^\mathrm{d}=-\epsilon k^2\mathbf{H\cdot \widehat{q}}$ with
\begin{equation}
\mathbf{H}=
\begin{pmatrix} \cos^2 \varphi +\kappa\sin^2 \varphi &\tf12(1-\kappa)\sin 2\varphi \\ 
 \tf12(1-\kappa)\sin 2\varphi & \sin^2 \varphi+ \kappa\cos^2 \varphi\end{pmatrix},
 \label{Hq} \end{equation}
 where $\kappa=K_2/K_1$. The the passive stress is linearized and Fourier-transformed as 
\begin{equation}
\widehat{\sigma}^\mathrm{p}_{ij}=\tf12 \chi (q^0_i \widehat{h}_j+\widehat{h}_iq^0_j)
-\tf14 (q^0_i\widehat{h}_j-\widehat{h}_iq^0_j) .
\label{passq}\end{equation}
After taking the curl of \eqref{nsg}, the contribution of the passive stress to the Fourier-transformed hydrodynamic equation is expressed as $\varepsilon_{ik}\partial_k\partial_j\widehat{\sigma}^\mathrm{p}_{ij}= - k^2\mathbf{f}\cdot \widehat{\mathbf{q}}$, where
\begin{equation}
\mathbf{f}=\begin{pmatrix}\chi\sin 2\varphi \\0\end{pmatrix}+
\frac{k^2}{8} \begin{pmatrix}[1+2\chi -\kappa(1-2\chi)]\sin 2\varphi \\ 
(1+2\chi)\sin^2 \varphi+ \kappa(1-2\chi\cos^2 \varphi \end{pmatrix}.
\label{passfk}\end{equation}
The  resulting form of the hydrodynamic equation is 
\begin{equation}
(\rho D_t  +\mu  k^2 )\widehat{\Psi}=-k^2 \mathbf{f}\cdot\widehat{q}.
 \label{qfl}\end{equation}

Turning to the dynamic equation of $\widehat{q}$, we compute the Fourier transform of \eqref{symasym}
\begin{equation}
\widehat{\mathbf{s}}= \tf12\epsilon k^2\widehat{\Psi} 
\begin{pmatrix}-\sin 2\varphi & \cos 2\varphi\\ \cos 2\varphi &\sin 2\varphi \end{pmatrix}, \quad
\widehat{\boldsymbol{\Omega}}=\tf 12\epsilon k^2\widehat{\Psi} \boldsymbol{\varepsilon},
\label{symasymc}\end{equation}
to obtain the transform of the linearized equation \eqref{qdyn} 
\begin{equation}
D_t \widehat{\mathbf{q}} = \widehat{\mathbf{h}}+\tf 12k^2\mathbf{g}\widehat{\Psi}, \quad 
\mathbf{g}=  \begin{pmatrix}\chi\sin 2\varphi \\ 
-(\tf 12 +\chi)\cos 2\varphi \end{pmatrix}. \label{qdyn2}\end{equation}

Stability requires the eigenvalues $\lambda$ of the following matrix to have negative real parts:
\begin{equation}
\mathbf{J}= \begin{pmatrix}-k^2\mu/\rho , & -k^2 f_1/\rho, & -k^2 f_2/\rho \\ 
\tf12 k^2g_1, & -2-k^2 H_{11}, & -k^2  H_{12}\\ \tf12 k^2g_2 ,&- k^2 H_{21}, &-k^2 H_{22}\end{pmatrix}.
\label{Ldyn} \end{equation}
The Routh--Hurwitz stability criterion is $a_0>0,\, a_1a_2-a_0>0$, where $a_j$ are the coefficients of $\mathrm{Det}( \lambda \mathbf{I} -\mathbf{J})=\sum_0^3 a_j \lambda^j$; \textbf{I} is the identity matrix. The leading terms of these conditions are computed as
\begin{align}
&a_0=2 k^4 (\mu/\rho)(\sin 2\varphi +\kappa\cos^2\varphi), \label{Ldet}\\
&a_1a_2-a_0=2 k^2[2\mu/\rho+1-\cos 2\varphi+ \kappa(1+\cos 2\varphi)],
\notag \end{align}
so that, unlike \eqref{jdet}, stability is ensured in the long-scale limit due to a stabilizing influence of perturbations of the modulus. However, the full expressions are quite complicated and do not formally exclude a short scale instability on wavelengths approaching the healing length. 

\begin{figure}[t]\centering
\includegraphics[width=0.4\textwidth]{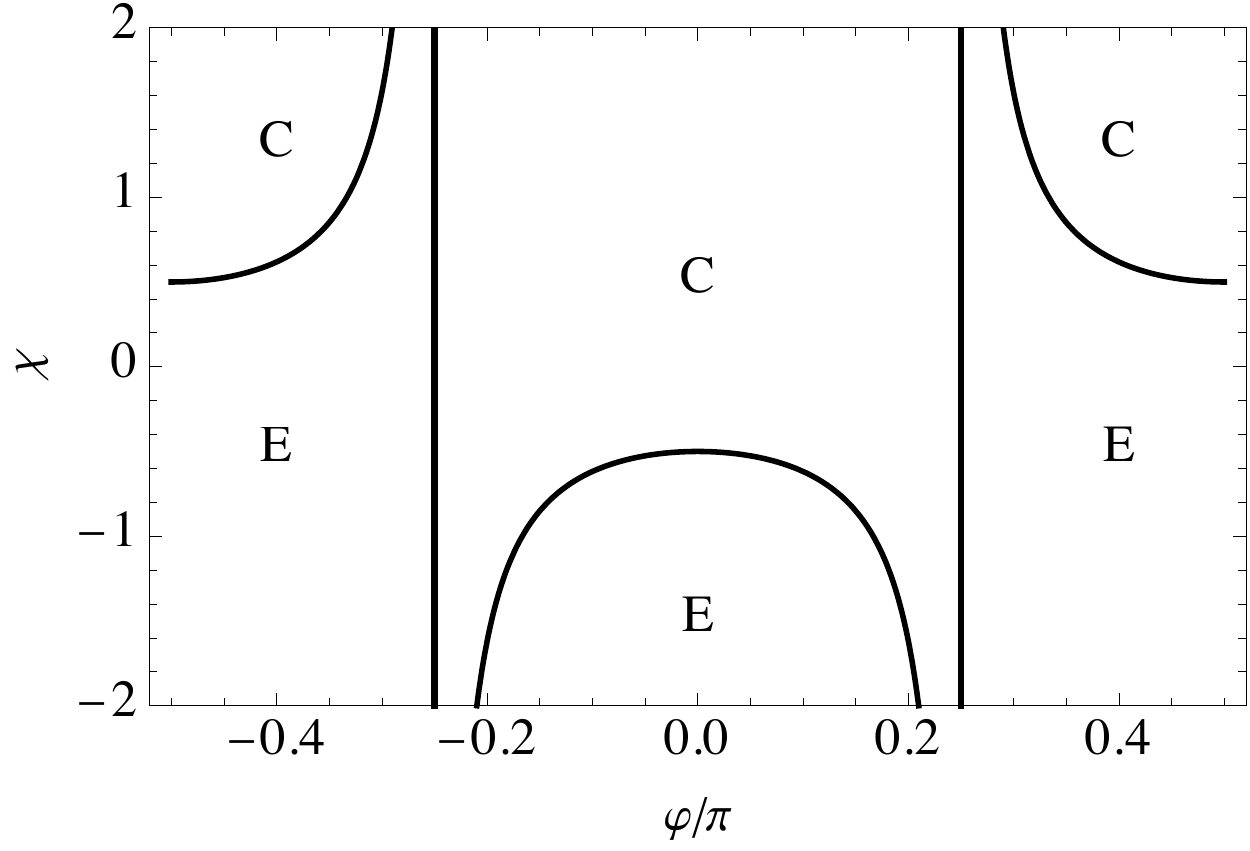}
\caption{Domains of instability due to contractile (C) and extensile (E) activity.}
\label{fqs}\end{figure}

Active stress $\sigma^\mathrm{a}=-\zeta Q_{ij}$ adds to the hydrodynamic equation \eqref{qfl} the term $\tf12k^2\zeta(\widehat{q}_1\sin 2\varphi -\widehat{q}_2 \cos 2\varphi)$, which modifies the entries in the first row of \eqref{Ldyn} to 
\begin{equation}
J_{12} = (\tf12\zeta\sin 2\varphi-k^2f_1)/\rho, \;\; J_{13} =-( \tf12\zeta\cos 2\varphi+k^2 f_2)/\rho.
\label{J123}\end{equation}
The threshold value $\zeta^*$ flipping the sign of $a_0$ satisfies
\begin{equation}
 zeta^*  \cos 2\varphi(1+2\chi\cos 2\varphi) =16\mu( \sin^2\varphi +\kappa\cos^2\varphi) .
\label{sp1}
\end{equation}
The right-hand side is always positive, and the left-hand side flips its sign at $\cos2\varphi=-\tf12\chi^{-1}$ and $\varphi=\pm \pi/4$, so that the plane $\varphi, \chi$ is separated into the regions of positive (C) and negative (E) thresholds shown in Fig.~\ref{fqs}. Since at any $\chi$ there is a perturbation angle falling in either region, activity of either sign can cause an instability: contractile in the region C and extensile in the region E, but in each case a finite strength is necessary, attained with respect to perturbations directed at the angle $\varphi$ that corresponds to the maximum absolute value of the expression multiplying $\zeta^*$, which, as can be easily seen, is attained at $\varphi=0,\pi$ for $\chi>0$ and at $\varphi=\pi/2$ for $\chi<0$, and for both cases equals $1+|\chi|$. Other extrema of this expression are lower by their absolute value, and instability is never observed at an arbitrarily small activity. 

Instability is always monotonic, since the leading term of the second Routh--Hurwitz stability criterion,
\begin{equation}
a_1a_2-a_0=2 [2\mu/\rho +1-\cos 2\varphi +\kappa(1+2\cos 2\varphi) ].
\label{sp23}
\end{equation}
is independent of activity and always positive.  

\subsection{Frictional motion}

A quasi-2D formulation of a problem in the 3D world is possible in a thin layer $0<z<l$ with tangential alignment on the upper and lower boundaries. The nematic alignment is then constant across a thin layer, but velocity is constant only in the case of free-slip boundary conditions. They are often assumed both in analysis and simulations, but may be realized only in a suspended film, which is known to be unstable. 

In the realistic case of solid confining walls, no-slip boundary conditions apply, leading to the parabolic velocity profile $\mathbf{v}(z)=\tf16\mathbf{v}_0 l^{-2}z(l-z)$, where $\mathbf{v}_0$ is the average in-plane velocity. The momentum flux to the walls $-2\mu \partial_z v_{|z=0}=\tf13\mu v_0 l^{-1}$ is compensated by the force $\mathbf{F}=-\nabla\cdot (p+\boldsymbol{\sigma})$ driving fluid's motion. This defines the average velocity $\mathbf{v}_0=3\mathbf{F}l/\mu$, and the total flux is $\mathbf{j}= \mathbf{F}l^2/\mu'$, which is equivalent, neglecting viscous anisotropy, to Darcy's law with the coefficient $3l^2/\mu$. Accordingly, \eqref{nsg} is replaced by the frictional equation
\begin{equation}
 \rho D_tv_i= - \mu' v_i - \partial_ip  - \partial_j \sigma^\mathrm{p}_{ij} .
   \label{kinsf}\end{equation}
containing the effective friction coefficient $\mu'=\tf13\mu/l^2$. We further restrict to this expression, which corresponds to the diagonal viscous stress tensor $\sigma_{ij}^\mathrm{s}=-\tf12\mu'|v|^2$. The  stress tensor accounting for anisotropy and modulus dependence is 
\begin{equation}
\sigma_{ij}^\mathrm{s}=-\tf12|v|^2 [(\mu'+\mu'_1\varrho)\delta_{ij }+\mu'_1Q_{ij}].
   \label{kinsf}\end{equation}

Frictional dynamics with a friction coefficient dependent on properties of particles and a substrate is also observed in an adsorbed nematic layer. In fluid layers, frictional dynamics restricts Galilean invariance, since non-uniform advection due to the parabolic velocity profile, may cause distortion of the nematic alignment field, which is only eliminated in the case of a high rotational mobility, $\Gamma\mu \gg 1$.  

In the case of frictional dynamics, the Fourier transform of \eqref{kinsf} replacing \eqref{qfl} is
\begin{equation}
 (\rho D_t  +\mu' )\widehat{\Psi}= \mathbf{f}\cdot\widehat{q},
 \label{qflf}\end{equation}
so that the elements of the first row of the matrix \eqref{Ldyn} supplemented by the active term become 
\begin{align}
J_{11}=\mu'/\rho, \quad & J_{12} = (f_1+\tf12\zeta\sin 2\varphi)/\rho, \cr
& J_{13} =(f_2- \tf12\zeta\cos 2\varphi)/\rho.
\label{J123f}\end{align}
Friction plays a stabilizing role, and activity fails to cause a long-scale monotonic instability in this case.

\section{Conclusions}

The vector-based description of two-dimensional dynamics adjusts the director-based theory to problems involving changes of the modulus of the order parameter. It not only enables easier computation, compared to commonly used routines based on flattening the 3D tensor representation, but evades degeneracy of elastic constants in the lowest order Landau-de Gennes description. 

The parallel stability analysis based on the director- and vector-based formalisms for a 2D slice of 3D space, which is possible due to the circular symmetry with respect to the unperturbed alignment, demonstrates inadequacy of the director-based description allowing for instability even in the absence of activity at a sufficiently low ratio of hydrodynamic to orientational viscosities, supporting the criticism of the standard theory \cite{bp}. Stability is restored on long scales due to perturbations of the modulus, which play a stabilizing role, allowing for instability only at a finite strength of either extensile or contractile activity. Instabilities are further suppressed in Hele-Shaw geometry with tangential nematic alignment on containing walls.

\end{document}